\documentclass[aip,pop,amsmath,amssymb,reprint]{revtex4-1}
\usepackage[T1]{fontenc} 
\usepackage{graphicx}
\usepackage{dcolumn}
\usepackage{bm}
\usepackage{comment}
\usepackage{soul}
\usepackage[dvipsnames]{xcolor}
\usepackage{color}

\def\nablav{\mbox{\boldmath$\nabla$}}

\def\div{\nablav\cdot}

\begin{document}

\preprint{}

\title
{Surface plasmons in  superintense laser-solid interactions}

\author{A.Macchi}
\affiliation{National Institute of Optics, National Research Council (CNR/INO),  Adriano Gozzini laboratory, via Giuseppe Moruzzi 1, 56124 Pisa, Italy}\email{andrea.macchi@ino.cnr.it}
\affiliation{Enrico Fermi Department of Physics, University of Pisa, largo Bruno Pontecorvo 3, 56127 Pisa, Italy}

\date{\today}

\begin{abstract}
We review studies of superintense laser interaction with solid targets where the generation of propagating surface plasmons (or surface waves) plays a key role. These studies include the onset of plasma instabilities at the irradiated surface, the enhancement of secondary emissions (protons, electrons, and photons as high harmonics in the XUV range) in femtosecond interactions with grating targets, and the generation of unipolar current pulses with picosecond duration. The experimental results give evidence of the existence of surface plasmons in the nonlinear regime of relativistic electron dynamics. These findings open up a route to the improvement of ultrashort laser-driven sources of energetic radiation and, more in general, to the extension of plasmonics in a high field regime.   
\end{abstract}

\maketitle

\section{Introduction}

The interaction of sub-picosecond, high intensity laser pulses with solid targets is the basis for the generation of ultrashort pulses of energetic radiation (including ions\cite{daidoRPP12,macchiRMP13}, electrons\cite{bastianiPRE97,caiPoP03,brandlPRL09,mordovanakisPRL09,wangPoP10,thevenetNP15} and photons\cite{tothPoP07,chenPRL08,teubnerRMP09,levyAPL10,thauryJPB10,zhangOE11}). The use of solid targets is complementary to that of gaseous ones, with advantages depending of the desired source characteristics and foreseen applications. Already at intensities $I>10^{16}~{\rm W cm}^{-2}$ the laser field is strong enough to cause instantaneous ionization, freeing the outer electrons within half a cycle. In addition, electron heating rises the mean electron energy well above the Fermi level while the collision frequency drops down in the ``skin'' layer where the field penetrates. Thus, any target material may be considered (at least to a first degree of approximation) as a collisionless classical plasma described by the dielectric function of a simple metal
\begin{equation}
\varepsilon(\omega)=1-\frac{\omega_p^2}{\omega^2}=1-\frac{n_e}{n_c(\omega)} \; ,
\label{eq:dielfuncLIN}
\end{equation}
where $n_e$ is the electron density, $\omega_p=(4\pi n_ee^2/m_e)^{1/2}$ is the plasma frequency, and $n_c=n_c(\omega)=m_e\omega^2/4\pi e^2$ is the cut-off (also named ``critical'') density for the frequency $\omega$.   

When the laser frequency $\omega_L$ is in the typical range for short pulse lasers, $\omega_p\gg\omega_L$ holds since $n_e \sim 10^2n_c$ for solid materials.
In addition, for ultrashort pulses (duration $<100$~fs) during the interaction the plasma expansion may be negligible (and further inhibited by the intense pressure of laser light) so that the target keeps a sharp density profile with a scalelength $L\ll\lambda_L$ where $\lambda_L=2\pi c/\omega_L$ is the laser wavelength.
In these conditions, the dielectric function jumps abruptly from unity to negative values $<-1$ across the vacuum-target interface, which allows the existence of propagating surface plasmons (SPs), also commonly referred to as surface waves or polaritons in solid state physics. The SP wavevector ${\bf k}_{\rm SP}$ is parallel to the surface and is related to the SP frequency $\omega_{\rm SP}$ by the dispersion relation (see e.g. Ref. \onlinecite[sec.2.2]{bookMaier2007} or Ref. \onlinecite[sec.68]{landau8})
\begin{equation}
\left(\frac{k_{\mbox{\tiny SP}}c}{\omega_{\mbox{\tiny SP}}}\right)^2=\frac{\varepsilon(\omega_{\mbox{\tiny SP}})}{\varepsilon(\omega_{\mbox{\tiny SP}})+1} \; .
\label{eq:SWdisp}
\end{equation}
SPs offer the possibility of resonant coupling with electromagnetic EM radiation and to confine the EM energy within a narrow region of sub-wavelength depth, since the SP field is evanescent on both the vacuum and material sides. These properties make SPs a building block of plasmonics with several applications (see e.g. Refs.\cite{barnesN03,maierJAP05,ozbayS06,plasmonics-resurrection}).    

Exploiting the properties of SPs in the high field regime, i.e. for field strengths of the order of those presently available with multi-terawatt and petawatt laser systems, is of interest as a way to advance laser-driven sources and other possible applications which may be inspired by plasmonics at low fields. However, such perspective has to deal with both theoretical and experimental issues. On the theory side, the properties of SPs in a regime where the electron dynamics is nonlinear and strongly relativistic are not well known. On the experiment side, plasmonics is tightly related to target nanostructuring, which for instance is necessary to couple laser light with SPs (see Section~\ref{sec:gratingtheory}). Obviously, nanostructures cannot survive for a long time as the material is transformed into a hot plasma. As it will be discussed in Section~\ref{sec:gratingexpts}, the coupling of high intensity laser pulses with structured targets requires the use of ultrashort (sub-picosecond), ``high contrast'' pulses. 

In this review paper, we will describe our contributions to the exploration of SPs at high fields along three lines of research. We try to keep the presentation at a non-specialist level accessible to a broad audience, beyond of the laser-plasma interaction community. The presentation of our contributions is preceded by a brief overview of SP theory (section \ref{sec:theory}) where open issues of the nonlinear high field regime are pointed out. 

Section~\ref{sec:instabilities} will be devoted to theory and simulation studies of laser-stimulated plasma instabilities at the target surface, where the role of SPs emerged as crucial. The earliest of these studies provided numerical evidence of SPs in the relativistic regime. 

In Section~\ref{sec:gratingexpts}, we will review a series of recent experiments on short pulse, high contrast laser interactions with solid grating targets, aimed at characterizing the effect of SPs on ``secondary'' emissions. The latter include protons, electrons and photons as XUV high harmonics of the incident laser. In turn, these experiments gave the main experimental evidence of ``relativistic'' SPs so far.

Finally, in Section~\ref{sec:unipolar} we will review a different series of experiments where SPs of picosecond duration are excited by the transient charge separation generated in intense laser-solid interactions. In this regime, SPs have possible applications in advanced devices for laser-driven ion acceleration and generation of intense EM pulses in the terahertz range.

\section{Theoretical considerations}
\label{sec:theory}

\subsection{Basic linear theory and issues for high fields}

In this section we briefly review the well-known elementary theory of SPs and a vacuum-plasma interface and we point out issues related to an extension of the theory for high field amplitudes and in the framework of intense laser-solid interactions (basics of the latter field may be found in textbooks, e.g. Refs.\onlinecite{gibbon-book,mulser-book,macchi-book}).

Let us consider a SP propagating along the boundary between vacuum and a simple metal or ``cold'' plasma. We assume the electron density profile to be $n_e=n_0\Theta(x)$ (i.e. the ``vacuum'' region is $x<0$), with $\omega_p^2=4\pi e^2n_0/m_e$,  and the SP to be a monochromatic plane wave propagating in the $y$ direction. Let the expressions for the EM fields of the SP have the form $f(x,y,t)=\tilde{f}(x){\rm e}^{iky-i\omega t}$. The components of the SP fields are found to be
\begin{eqnarray}
\tilde{E}_y(x)&=&E_0\left[\Theta(-x){\rm e}^{+q_{<}x}+\Theta(+x){\rm e}^{-q_{>}x}\right]  \;, \\
\tilde{B}_z(x) 
&=& \frac{i\omega /c}{q_{<}}
E_0\left[\Theta(-x){\rm e}^{+q_{<}x}
 +\Theta(+x){\rm e}^{-q_{>}x}\right]  ,
\\
\tilde{E}_x(x) 
&=& -ik E_0 \left[\Theta(-x)\frac{{\rm e}^{+q_{<}x}}{q_{<}}
                              -\Theta(+x)\frac{{\rm e}^{-q_{>}x}}{q_{>}} \right] ,
\label{eq:SW_fields}
\end{eqnarray}
where, posing $\omega_p^2/\omega^2 \equiv \alpha$ for brevity, 
\begin{eqnarray}
q_{>}&=&({\omega}/{c})({\alpha-1})/(\alpha-2)^{1/2}=(\alpha-1)^{1/2}k \; , \\
q_{<}&=&({\omega}/{c})({1}/(\alpha-2)^{1/2}=(\alpha-1)^{-1/2}k \; ,
\end{eqnarray}
so that, using the dispersion relation (\ref{eq:SWdisp}), 
$(\alpha-1)q_{<}=q_{>}$ and $q_{<}q_{>}=k^2$. 

\begin{figure}
\includegraphics[width=0.45\textwidth]{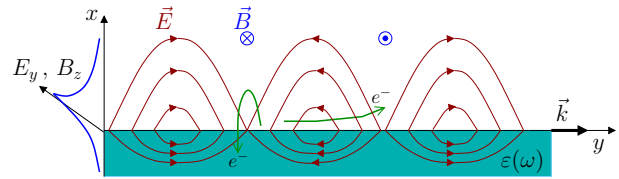}
\caption{Sketch of the EM fields of a surface plasmon propagating in the $y$ direction along the interface between vacuum and a step-boundary plasma having a value of the dielectric function $\varepsilon<-1$. Also sketched are examples of the trajectories of electrons accelerated both across the interface by the transverse field $E_x$ and along the surface by the longitudinal field $E_y$. \label{fig:SWcartoon}}
\end{figure}

A sketch of the SP field is shown in Fig.\ref{fig:SWcartoon}.
Notice that $E_y$ and $B_z$ are continuous at $x=0$ while $E_x$ is discontinuous, so there is a surface charge density $\sigma(y,t)=\tilde{\sigma}{\rm e}^{iky-i\omega t}$ such that $\tilde{\sigma}=4\pi[\tilde{E}_x(0^{+})-\tilde{E}_x(0^{-})]$. No volume charge density exists since $\div{\bf E}=0$ as can be easily verified. This means that in the linear model electrons not allowed to enter the vacuum region, despite the presence of a finite $E_x$. If a finite temperature plasma is assumed\cite{kawPF70}, the surface charge is replaced by a thin layer of charge separation having a depth of the order of the Debye length $\lambda_D=v_{th}/\omega_p$ (where $v_{th}$ is the thermal velocity) but a reflecting boundary condition for the electron velocity at the $x=0$ plane is still used. Such assumption is necessary in a linear theory since the equations of motion cannot be linearized for electrons entering the vacuum side. From a physical point of view, the reflecting boundary mimics a thin sheath region with an electric potential barrier preventing thermal electrons from escaping at the surface. Thus, in the presence of a driving field $E_x \simeq \bar{E}_x{\rm e}^{-i\omega t}$ the reflecting boundary condition is appropriate only as long as the oscillation velocity $v_x<v_{th}$, with $v_x \sim e\bar{E}_x/m_e\omega$. However, for high fields such that $v_x>v_{th}$ the boundary condition becomes inappropriate. The situation is analogous to the modeling of absorption of an EM wave incident on the surface of the step-boundary plasma which for low fields can be described by Fresnel-like formulas with an absorption coefficient determined by processes, such as the anomalous skin effect or sheath inverse Bremsstrahlung, which occur in the skin layer \cite{weibelPoF67,cattoPoF77,rozmusPoP96,yangPoP96}, while for high fields the nonlinear motion of electrons across the interface may lead to ``vacuum heating'' absorption \cite{brunelPRL87,brunelPF88} (see also \onlinecite[p.161]{gibbon-book}) and ``fast'' electron generation. Thus, one may expect nonlinear kinetic effects to cause damping of high-amplitude SPs. 

Another possible limiting factor to the amplitude of a SPs comes from the condition that the oscillation velocity along the propagation direction ($v_y$) must be less than the phase velocity $v_p=\omega/k$, analogous to the ``wavebreaking'' limit for longitudinal plasma waves (or bulk plasmons). In the limit $\omega\ll\omega_p$, we have $v_p \simeq c$ so that in the linear regime a SP remains far from wavebreaking. The situation, however, may be different for SPs of relativistic amplitude for which $v_y$ approaches $c$. 

As field amplitudes reach values of the order of $m_ec\omega/e$, the electron dynamics becomes relativistic. For a laser pulse, this occurs when the dimensionless parameter
\begin{equation}
a_L \equiv \frac{eE_L}{m_e\omega c}=\left(\frac{I_L\lambda_L^2}{10^{18}~{\rm W cm}^{-2}\mu{\rm m}^2}\right)^{1/2}
\label{eq:a_L}
\end{equation}
attains values $a_L \gtrsim 1$. In this so-called ``relativistic'' regime the response of the medium is nonlinear, i.e. the current density is not linear in the field amplitude. The propagation of an EM wave in an homogeneous plasma can be approximately described by a nonlinear dielectric function
\begin{equation}
\varepsilon_{\rm NL}(\omega)=1-\frac{\omega_p^2}{\gamma_e\omega^2} \; ,
\label{eq:dielfuncNL}
\end{equation}
where $\gamma_e$ is the relativistic factor of electrons, which depends on the EM field amplitudes; as an example, $\gamma_e=(1+a_L^2/2)^{1/2}$ for an electron in a plane wave. This is equivalent to assume an effective mass $m_{\rm eff}=m_e\gamma_e$ in the plasma frequency, which accounts for the relativistic inertia due to the oscillatory motion. Since in a nonlinear regime the response of the medium depends on the type of wave, Eq.(\ref{eq:dielfuncNL}) cannot be used straightforwardly to obtain a relativistic generalization of the SP dispersion relation\cite{liuPoP15} (\ref{eq:SWdisp}). At most, the effective mass concept may be used for qualitative hints.

\subsection{Electron heating and acceleration}
\label{sec:elecaccel}

SP enhancement of photoelectron emission is widely investigated in plasmonics for, e.g., the development of efficient ultrafast photocathodes\cite{tsangPRB91,dombiOE08,hwangPRB09,raczAPL11,watanabeJAP11,liPRL13,gongPRApp14,onoJPD15}. In laser-grating interaction experiments at low intensity, anomalously ``hot'' photoelectrons were observed and attributed to ponderomotive acceleration from the evanescent SP field\cite{zawadzkaAPL01,kupersztychPRL01,kupersztychPoP04,irvinePRL04,irvinePRA06}. In the high field regime, the transfer of energy from SPs to electrons has peculiar properties.  

As already noticed above, electron oscillations driven across the plasma-vacuum interface by the transverse component of the SP electric field ($E_x$ in Fig.\ref{fig:SWcartoon}) may lead to damping of the SPs and, equivalently, to the absorption of EM energy by the electrons. Thus, SP excitation may be used to enhance heating of a solid target, which is the key for some applications including proton acceleration (see Section~\ref{sec:protons}).

Since the SP phase velocity $v_p<c$, the longitudinal electric field component ($E_y$ in Fig.\ref{fig:SWcartoon}) can lead to ``surfing'' acceleration of electrons along the direction of propagation. In particular, $v_p$ approaches $c$ for $\omega\ll\omega_p$, which makes SPs suitable for the acceleration of relativistic electrons. The situation is rather similar to the laser wakefield acceleration (LWFA) in homogeneous ``bulk'' plasmas \cite{tajimaPRL79}, and indeed to estimate the energy gain  one can use a similar model exploiting the fact that the SP field is electrostatic in a boosted reference frame moving at $v_p$. As an important difference with respect to LWFA, the rapidly evanescent transverse component ($E_x$ in Fig.\ref{fig:SWcartoon}) drives electrons out of the region if which the SP is localized, and such effect is a limiting factor for the energy gain. Related modeling and experiments are reported in Ref.\onlinecite{fedeliPRL16}. 

An important aspect of surfing acceleration in laser-driven SPs is the injections of electrons, i.e. the characterization and control of optimal initial conditions. This problem was theoretically investigated in Ref.\onlinecite{ricondaPoP15} which focuses on injection induced by the Lorentz force of the laser field. Another possibility is self-injection near the wavebreaking threshold in the relativistic SP regime.

\subsection{The coupling problem. Grating targets}
\label{sec:gratingtheory}

A key issue of coupling SPs with laser light is that phase matching (PM) between the SP and an incident EM wave is not possible at a flat interface. Assuming that the laser pulse is modeled as an EM plane wave incident at an angle $\theta$ on the target surface, the PM conditions imply the equations
\begin{equation}
  \omega_L=\omega_{SP} \; , \qquad \; k_{L,\parallel}=k_{\rm SP} \; ,
\label{eq:PM1}
\end{equation}
where ``$\parallel$'' indicates the wavevector component parallel to the surface. Now, $k_{L,\parallel}=(\omega_L/c)\sin\theta$ while $k_{{\rm SP},\parallel}>\omega_{\rm SP}/c$, so that Eqs.(\ref{eq:PM1}) have no solution. 

In standard plasmonics, the coupling issue can be tackled efficiently by different approaches\cite{bookMaier2007}. The only approach to which we will refer in this context is to replace the flat surface with a periodically engraved one, i.e. a grating. This approach is suitable for high fields since it does not need the laser pulse to propagate across a transparent medium (as e.g. in ``prism coupling'') which would be rapidly ionized by the laser field. 

In a periodic medium with lattice spacing $d$, the Floquet-Bloch theorem implies that the dispersion relation is replicated as a function of the wavevector with periodicity $q=2\pi/d$. This is equivalent to assume that the dispersion relation is folded into the Brillouin zone $|k_{\parallel}|<\pi/d$ or to modify Eqs.(\ref{eq:PM1}) as follows,
\begin{equation}
  \omega_L=\omega_{SP} \; ,  \qquad \; k_{L,\parallel}=k_{\rm SP}+nq \; ,
\label{eq:PM2}
\end{equation}
with $n$ an integer number. For fixed values of $\omega_L$, $d$ and $n$ Eqs.(\ref{eq:PM2}) are satisfied for a well defined value of $\theta$. If $\omega_p\ll\omega_L$, the resonant condition is approximately given by 
\begin{equation}
\sin\theta=n\frac{\lambda}{d}-1 \; .
\label{eq:resangle}
\end{equation}
Usually, a grating is designed to yield resonant coupling for $n=1$.
Notice that Eq.(\ref{eq:resangle}) is equivalent to the condition that the maximum of light diffracted from the grating at the $n$-th order occurs at $-90^{\circ}$, i.e. along the surface. 

\begin{figure}[t]
\includegraphics[width=0.45\textwidth]{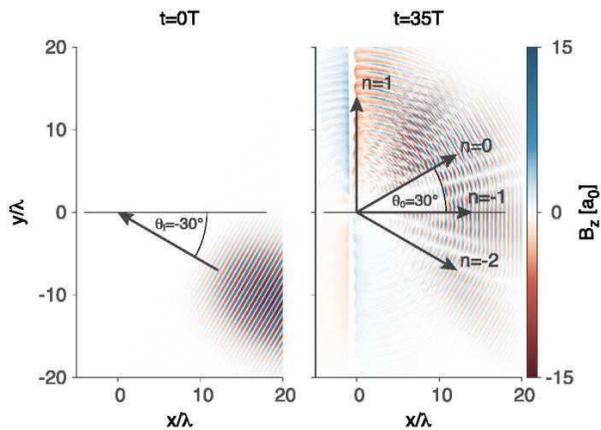}
\caption{Two-dimensional PIC simulation of intense short pulse interaction with a grating target at the resonant angle $\theta_{\rm r}$ for SP excitation\cite{fedeliAPL17}. A short pulse with peak amplitude $a_L=15$, waist size of $5\lambda_L$, duration of $12\lambda_L/c$ and $P$-polarization impinges on a plasma target of electrons density $n_e=128n_c$ and a sinusoidal surface modulation with pitch $d=2\lambda_L$, corresponding to $\theta_{\rm r}=30^{\circ}$, and depth $0.25\lambda_L$.  The contourplot of $B_z$, the magnetic field component perpendicular to the simulation plane, at two different times show the diffraction of the incident light at several orders. The $n=1$ order corresponds to propagation along the surface, where intense localized fields are observed. Reprinted from Ref.\onlinecite{fedeliAPL17}. Used by permission.
\label{fig:diffraction}}
\end{figure}

As an example, Fig.\ref{fig:diffraction} shows results from a particle-in-cell (PIC) simulation of high intensity laser interaction with a grating\cite{fedeliAPL17} at the resonant angle for SP excitation. The diffraction of light at several order is observed, with $n=1$ corresponding to the direction tangent to the surface, along which intense localized fields are observed. 

In principle, the dispersion relation is affected by the surface modulation, but the corrections turn out to be of the order of $(\delta/\lambda_L)^2$ where $\delta$ is the grating depth, and are thus usually negligible for shallow gratings such that $\delta\ll\lambda_L$. Issues related to the exploitation of grating coupling in the high field regime will be discussed in Section~\ref{sec:gratingexpts}.

Going beyond the linear regime, a nonlinear conversion of the laser pulse into a pair of SPs is possible also at a plane interface, for instance by a three-wave process in which two SPs are excited by a pump wave\cite{gradovPP80,leePRE99,numarPoP07,akimovPoP07,akimovPoP08}. We named this process ``two surface-wave decay''\cite{macchiPoP02} although ``two surface-plasmon decay'' (TSPD) would be more adequate. In the case the two SPs are excited by the oscillating electric field of the laser pulse, the nonlinear matching conditions are
\begin{equation}
  \omega_L=\omega_{{\rm SP},1}+\omega_{{\rm SP},2}+ \; , \qquad \; k_{L,\parallel}=k_{{\rm SP},1}+k_{{\rm SP, 2}} \; ,
\label{eq:PMNL}
\end{equation}
where $\omega_{SP,i}$ and $k_{{\rm SP},i}$ ($i=1,2$) are the frequencies and wavevectors of the two SPs. At a plane interface, the electric field needs to have a component perpendicular to the surface in order to drive the electron oscillations, thus the process is possible for oblique incidence and $P$-polarization.

In Section~\ref{sec:paramosc}, we describe a particular version of TSPD \cite{macchiPRL01,macchiPoP02} where the SPs are driven by the magnetic force term of the laser pulse. In this case the SP excitation is allowed also at normal incidence and for $S$-polarization.

\section{Surface plasmon impact on plasma instabilities}
\label{sec:instabilities}

\subsection{Parametric excitation of surface plasmons}
\label{sec:paramosc}

\begin{figure}
\includegraphics[width=0.45\textwidth]{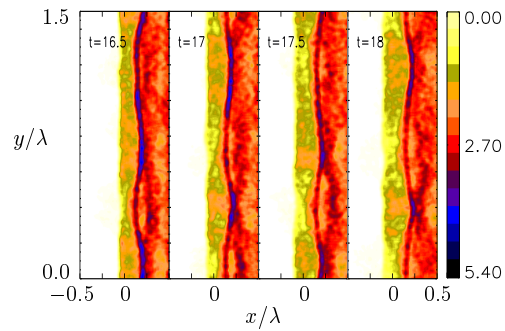}
\caption{Parametrically excited surface oscillations in a 2D simulation of laser interaction with an overdense plasma\cite{macchiPRL01}. A plane wave pulse with amplitude $a_L=0.85$ and $S$-polarization (i.e. normal to the simulation plane) is normally incident from the left side on a plasma slab with density $n_e=3n_c$. Snapshots of the electron density are shown at four times differing by a laser halfcycle to give evidence of a surface oscillation at the laser frequency $\omega_L$. The oscillation is generated via the parametric decay of the plane oscillation driven by the ${\bf v}\times{\bf B}$ force at twice the laser frequency ($2\omega_L$) into a standing surface plasmon at half-frequency $\omega_L$. Reprinted from Ref.\onlinecite{macchiPRL01}. Used by permission. \label{fig:TSPD}}
\end{figure}

Early studies of the interaction of superintense laser pulses with high density, step-boundary plasmas\cite{wilksPRL92} showed the occurrence of surface rippling, which was also believed to cause a transition from specular to diffuse reflection in some experiments\cite{norreysPRL95}.
Within a campaign oriented to understand the origin of surface rippling, two-dimensional (2D),plane wave PIC simulations showed the onset of ripples oscillating at the frequency $\omega_L$ of the driving laser (Fig.\ref{fig:TSPD})\cite{macchiPRL01}, which were interpreted as a standing, nonlinear surface wave originating from a parametric process. At normal incidence, the magnetic (${\bf v}\times{\bf B}$) force drives a plane, sweeping oscillation of the surface at the frequency $2\omega_L$. Such oscillation can couple to a pair of counterpropagating SPs both of frequency $\omega_{\rm SP}=\omega_L$ and opposite wavevectors $\pm{\bf k}_{\rm SP}$. This particular TSPD process can be considered as an EM version of Faraday wave (or Faraday ripple) generation, a classic example of parametric resonance in hydrodynamics. 

A non-relativistic analytic calculation of the growth rate of the ``$2\omega_L\rightarrow\omega_L+\omega_L$'' process was first presented in Ref.\onlinecite{macchiPoP02} and later refined\cite{macchiAPB04} in order to account for the effect of electron temperature and surface charges. The PIC simulations showed that the wavelength of the standing SP was smaller than the prediction of the linear dispersion relation (\ref{eq:SWdisp}) and decreased with growing laser intensity. Qualitatively, these effects might be attributed to a nonlinear decrease of the plasma frequency, i.e. to an effective mass as in Eq.(\ref{eq:dielfuncNL}) due to the laser-driven oscillation in the direction normal to the simulation plane, which overlaps to the motion due to the SP field. Little may be thus inferred about the correct dispersion relation of SPs having ``relativistic'' amplitude.
  
In Ref.\onlinecite{macchiAPB04} it was also shown that TSPD in laser-solid interactions may lead to a localized enhancement of ``vacuum heating'' of electrons, causing the imprint of the SP wavelength on the electron distribution. This effect could play a role in the formation of ``fast'' electron filaments as observed in some simulations\cite{lasinskiPoP99,sentokuPoP00}.  A more recent simulation study\cite{klugePoP15} also suggested that a TSPD process may seed electron filamentation, modulations in accelerated ions (see Section~\ref{sec:protons}) and Rayleigh-Taylor-like rippling instabilities (see Section~\ref{sec:RTI}). 
Another study has addressed the impact of TSPD on high harmonic generation (see Section~\ref{sec:harmonics}) from the surface \cite{anderbruggePRL12}.

\subsection{Plasmonic enhancement of Rayleigh-Taylor instabilities}
\label{sec:RTI}

The issue of understanding the origin of surface ripples was further stimulated by the interest in radiation pressure acceleration \cite{esirkepovPRL04,macchiPRL05,klimoPRSTAB08,robinsonNJP08,chenPoP11,khudikPoP14} where the rippling instability may cause disruption of the target and affect the spatial quality of the accelerated ions\cite{palmerPRL12}. In the case of thin foil targets, a Rayleigh-Taylor-type instability (RTI)\cite{ottPRL72,pegoraroPRL07}  has been considered as the most likely mechanism for the rippling onset, but purely hydrodynamic RTI models  did not explain the spatial scale of the unstable mode, which was of the order of the laser wavelength $\lambda_L$ as observed in simulations. 

The explanation proposed independently in Refs.\onlinecite{sgattoniPRE15,eliassonNJP15} is based on the coupling of the laser field with a ripple perturbation. For a 2D grating-like ripple, the electric field component perpendicular to the grooves is strongly enhanced inside the ripple valleys with respect to its value at a plane surface, so that the local radiation pressure enforces the perturbation. For normal laser incidence, Eq.\ref{eq:resangle} implies a SP resonance if $\lambda_L \simeq d$, which maximizes the field enhancement and produces a strong modulation of the radiation pressure on a scale $\sim\lambda_L$ and gives a seed for the RTI. 

\begin{figure}
\includegraphics[width=0.45\textwidth]{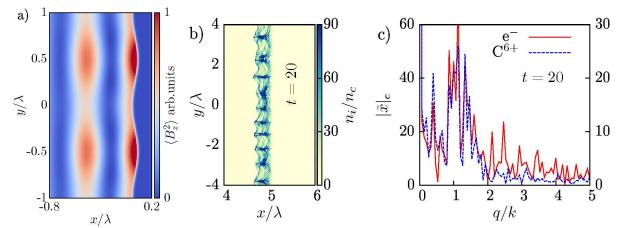}
\caption{a): 2D simulation of the interaction of a plane wave of wavelength $\lambda_L$ with a sinusoidally rippled surface having pitch $d=\lambda_L$. The polarization is in the simulation plane. The map of the magnetic field ($B_z$) shows enhancement inside the ripple valleys. b): 2D PIC simulation showing the onset of surface rippling in the electron density. The simulation is for a Carbon target with density $n_e=37n_c$ and a plane wave pulse with amplitude $a_L=66$, normally incident from the left. c): the spatial Fourier spectrum of the density perturbation for both electrons and Carbon ions for the same simulation of~b), showing a peak for a wavevector $q \simeq 2\pi/\lambda_L$. Adapted from Ref.\onlinecite{sgattoniPRE15}. Used by permission.
\label{fig:SP-RTI}}
\end{figure}

Figure~\ref{fig:SP-RTI} shows results from 2D simulations\cite{sgattoniPRE15} where the the field enhancement in the ripple valleys and the growth of a RTI mode with wavelength $\sim\lambda_L$ is apparent. The ``plasmonic'' nature of the instability is also supported from the observation that the unstable mode is first observed in the electron density. 3D simulations show that the development of the instability is affected both by geometry and kinetic effects. In the case of circular polarization, which moderates the electron heating, the plasmonic-enhanced RTI develops a pattern of hexagonal structures \cite{sgattoniPRE15}.

\section{Plasmon-ehnanced emission in grating targets}
\label{sec:gratingexpts}

Attempts to exploit the resonant excitation of SPs in grating targets in order to achieve high absorption of intense femtosecond pulses already started more than two decades ago\cite{gauthierSPIE95}. However, until recently experiments\cite{kahalyPRL08,huPoP10,bagchiPoP12} were limited to relatively modest intensities $\lesssim 10^{16}~{\rm W cm}^{-2}$ because of the prepulses present in ultrashort, multi-terawatt laser system as either nanosecond pedestals or secondary femtosecond pulses preceding the main one with the highest intensity. The main pulse-to-prepulse intensity ratio or ``contrast'' was not high enough to prevent early ionization and plasma formation, destroying the shallow gratings before the intense interaction. More recently, the development of ionization shutters\cite{kapteynOL91,dromeyRSI04,levyOL07,thauryNP07}, commonly referred to as ``plasma mirrors'', has made possible to achieve contrast values $\geq 10^{12}$ (usually measured a few ps before the main pulse) so that in principle the prepulse should not cause any ionization in a dielectric material even at intensities exceeding $10^{18}~{\rm W cm}^{-2}$, i.e. in the relativistic regime. In such high contrast conditions, it has become meaningful to study superintense interactions with shallow gratings (following the input from simulations\cite{raynaudPoP07,bigongiariPoP11,bigongiariPoP13,blancoNJP17}) and more in general with targets structured on the sub-micrometric scale\cite{margaronePRL12,floquetJAP13,purvisNatPhot13,andreevPPCF15,JiSR16,daluiAIPADV17,khaganiSR17,luebckeSR17}.

Experiments performed at the SLIC facility of CEA Saclay (France) using the high-contrast ($>10^{12}$ pulse-to-prepulse intensity ratio) UHI laser have characterized three types of radiation emission (proton, electrons, and high harmonics) either enhanced or directly driven by SPs excited in grating targets. The main observations are summarized in the following. 

\subsection{Protons} 
\label{sec:protons} 

\begin{figure}[t]
\includegraphics[width=0.45\textwidth]{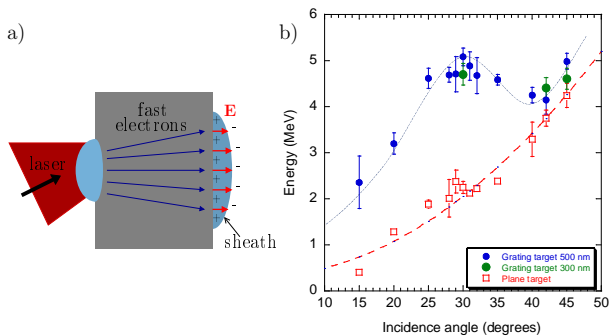}
\caption{Plasmon-enhanced TNSA of protons\cite{ceccottiPRL13}. a): schematic of TNSA. The fast electrons produced by the interaction at the front side cross the target and produce a sheath at the rear side, where ions are accelerated.  
b): experimental data from the interactions of a high-contrast $25$~fs, $2.5 \times 10^{19}~{\rm W cm}^{-2}$ laser pulse with solid plastic targets. The cut-off energy of protons emitted from the rear measured as a function of the incidence angle from both flat and grating targets (for two different values of the grating depth). An up to 2.5-fold energy increase is observed for gratings, with a broad maximum around the resonant angle for SP excitation ($30^{\circ}$). Data from Ref.\onlinecite{ceccottiPRL13}.
\label{fig:protons}}
\end{figure}

In solid targets irradiated at high intensity, protons are produced mainly via the target normal sheath acceleration (TNSA) mechanism\cite{wilksPoP01,macchiRMP13} (Fig.\ref{fig:protons}~a).
Briefly, the ``fast'' energetic electrons produced by the laser plasma interaction at the front side of the target cross the latter and produce a charged sheath at the rear side. The electric field in the sheath backholds electrons escaping in vacuum and accelerates ions, predominantly protons present either as a component of the target material or as impurities. The strength of the accelerating field is directly related to the fast electron temperature, so that an enhancement of the energy of the electrons which penetrate the target is expected to lead to more energetic protons. 

In the experiment at SLIC\cite{ceccottiPRL13}, the irradiation of grating targets produced a $\sim 2.5$-fold increase in the cut-off energy of the protons with respect to a flat target of the same material and thickness, at an angle of incidence equal to the value expected for SP excitation (Fig.\ref{fig:protons}~b). The measurements at different angles suggest a broad resonance, with a non-negligible contribution to heating enhancement by purely geometrical effects at small angles. 

\subsection{Electrons} 
\label{sec:electrons}

As discussed in section~\ref{sec:elecaccel}, electrons are not only accelerated inside the target by the transverse force of the SP, but also experience direct ``surfing'' acceleration along the SP propagation direction, driven by the longitudinal field component. 

\begin{figure}[b]
\includegraphics[width=0.45\textwidth]{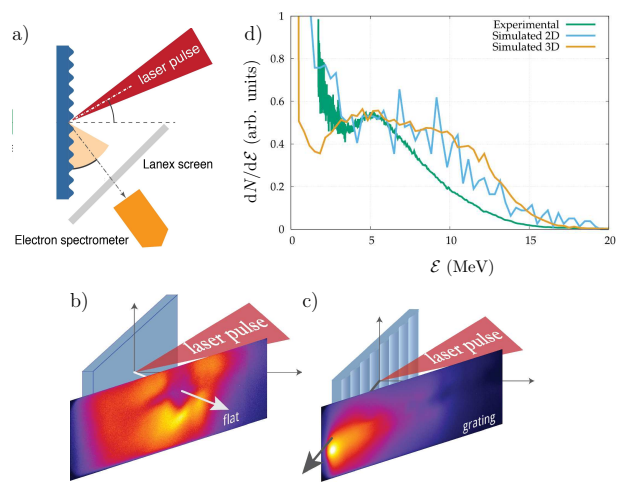}
\caption{Electron acceleration by surface plasmons\cite{fedeliPRL16}. a): basic experimental set-up. Laser and target parameters are the same as in Fig.\ref{fig:protons}. b): the image on the lanex screen for a flat target, showing a diffuse angular distribution of electrons. c): image for a grating target irradiated at the resonant angle for SP excitation, showing an highly collimated emission close to the tangent at the target surface. d): energy spectrum for the collimated electron emission, compared to 2D and 3D simulation results. Figures adapted from Ref.\onlinecite{fedeliPRL16} and Ref.\onlinecite{sgattoniPPCF16}. Used by permission.
\label{fig:electrons}}
\end{figure}

Fig.\ref{fig:electrons} summarizes the main results obtain in a devoted experiment at SLIC\cite{fedeliPRL16}. When a grating target is irradiated at the resonant angle for SP excitation, a collimated bunch of electrons is observed in a direction close to the surface. This is in striking contrast with the diffuse emission from a flat target. Electron spectra show that the collimated electrons have a non-thermal spectrum extending up to tens of MeV, while for flat targets the maximum electron energies always remained below the detector cut-off. The electron bunch contains a total charge of $\sim 100$~pC, which is of potential interest for applications. More recent measurements\cite{cantono-thesis} show that both the bunch charge and energy may be increased by the use of blazed gratings, i.e. having an asymmetric triangular profile.

PIC simulations closely reproduce the experimental results. These are also in qualitative agreement with the predictions of the simple model of acceleration by the SP\cite{fedeliPRL16}, which estimates the relativistic factor of electrons $\gamma_e$  and the emission angle $\theta_e$ (measured from the surface) as
\begin{equation}
\gamma_e \simeq 1+a_{\rm sp}\left(\frac{n_e}{n_c}\right) \; , \qquad
\theta_e \simeq \frac{1}{\gamma_e} \; ,
\label{eq:elecmodel}
\end{equation}
where $a_{\rm sp}$ is the dimensionless amplitude of the SP field. Simulations performed for $n_e/n_c=50$ showed that $a_{\rm SP} \simeq 1$, so that Eqs.\ref{eq:elecmodel} $\gamma_e \simeq 50$ and $\theta_e \simeq 8^{\circ}$, which are fairly consistent with the simulation results. It should be noticed that the density in the PIC simulations is limited by computational constraints to values lower than for a solid target, and that for realistic values $n_e>100n_c$ Eqs.\ref{eq:elecmodel} would predict much higher energies. However, achieving such energies in the experiment might be limited by the laser spot size, which is smaller than the required acceleration length\cite{fedeliPRL16} $L_{\rm acc} \simeq (\lambda/2\pi)(n_e/n_c)$. 

\subsection{High harmonics} 
\label{sec:harmonics}

Like fast electron generation, the emission of high harmonics (HH) of the incident laser pulse from solid targets (see Refs.\onlinecite{teubnerRMP09,thauryJPB10} for reviews) in the high-intensity regime is related to the nonlinear electron oscillations driven across the vacuum-target interface. Maybe the simplest model for HH generation is based on considering the collective oscillation as an oscillating mirror. An alternative mechanism is based on the radiating modes excited by the electrons returning in the target (``coherent wake emission''). In both cases, HH from a flat target are emitted in the direction of specular reflection, i.e. all harmonics are collinear. For applications, the angular separation of harmonics is desirable. To this aim, grating targets (in conditions far from SP resonance) have been used experimentally\cite{yeungOL11,cerchezPRL13} in order for each HH to be emitted at a particular angle determined by the diffraction grating equation. 

\begin{figure}[t]
\includegraphics[width=0.45\textwidth]{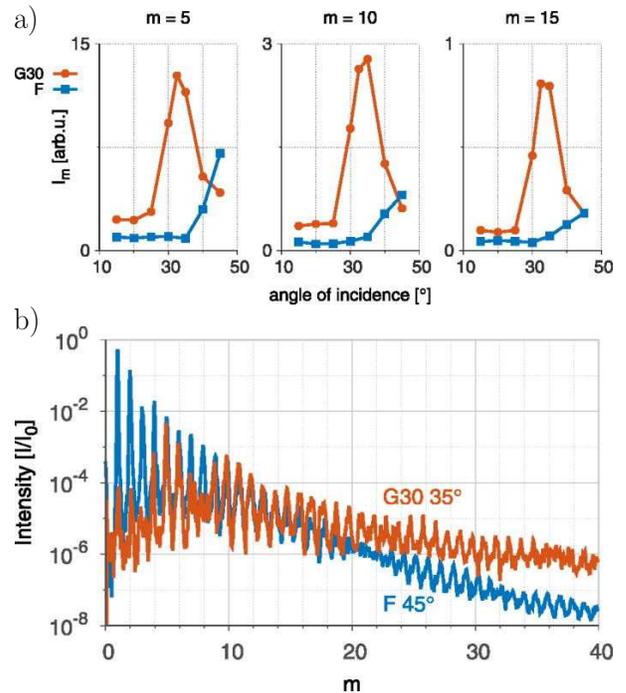}
\caption{Two-dimensional PIC simulation of plasmonic-enhanced high harmonic (HH) generation\cite{fedeliAPL17}. Laser and target parameters are the same as in Fig.\ref{fig:diffraction}. a): total (angle-integrated) yield for the harmonics of order $m=5,\,10,\, 15$ as a function of the incidence angle for flat targets (F, blue curves) and grating targets (G30, red curves) with pitch $d=2\lambda$. A sharp peak is observed for grating targets near the expected value of the resonant angle for SP excitation ($\theta=30^{\circ}$).  b): the comparison of harmonic spectra for the flat target in specular direction, and for the grating target in the nearly tangent direction. A strong enhancement for the highest harmonic orders is observed.
Reprinted from Ref.\onlinecite{fedeliAPL17}. Used by permission. 
\label{fig:harmonics}}
\end{figure}

The investigation of HH generation in gratings irradiated at the resonant angle for SP generation has the aim to combine the enhancement of the driving field at the surface with the angular separation. PIC simulations\cite{fedeliAPL17} have shown a significant plasmonic enhancement of the highest harmonics, with nearly two order of magnitude increase in the 40th harmonic (Fig.\ref{fig:harmonics}) at the SP resonance peak, and for HH emission in the direction close to the tangent at the target surface.  This suggests that the strongest emission may be related to nonlinear scattering by the accelerated electrons (section~\ref{sec:electrons}), which cross the SP fields near to the surface of the grating. The maximum HH yield actually occurs at an incidence angle $\theta=35^{\circ}$ slightly larger than the resonant value given by linear theory ($\theta_{\rm r}=30^{\circ}$), which is a possible indication of nonlinear effects in SP generation.

The numerical predictions have been substantially confirmed by recent measurements at SLIC\cite{cantono-thesis}. As an important experimental observation, the plasmonic enhancement of the HH signal has been observed also when a small preplasma of sub-wavelength scale was created by on the grating surface by using a short prepulse. The gain in HH intensity due to the preplasma is of the same order of that observed for flat targets\cite{kahalyPRL13}. This indicates that the grating periodicity was not washed out by the preplasma formation, despite the shallow depth of the grating, and that in general using suitable prepulses it is possible to optimize the laser-target coupling also in the presence of sub-micrometric structuring. In particular it may even be possible to create a transient grating in flat targets\cite{monchocePRL14}.
 
\section{Unipolar waves and applications}
\label {sec:unipolar}

The electric field generated by hot electrons in vacuum during the TNSA process (Fig.\ref{fig:protons}~a) is electrostatic only in the 1D approximation. Since the sheath is limited in the transverse direction (perpendicular to the beam expansion direction), the transient charge separation acts as a dipole antenna and can generate EM waves. This phenomenon has been investigated as a possible source of THz radiation \cite{gopalPRL13,tokitaSR15,poyePRE15}, because the duration of the transient stage is of the order of a few picoseconds which corresponds to some $\sim 10^{11}-10^{12}~\mbox{s}^{-1}$ frequencies. 

\begin{figure}[t]
\begin{center}
\includegraphics[width=0.45\textwidth]{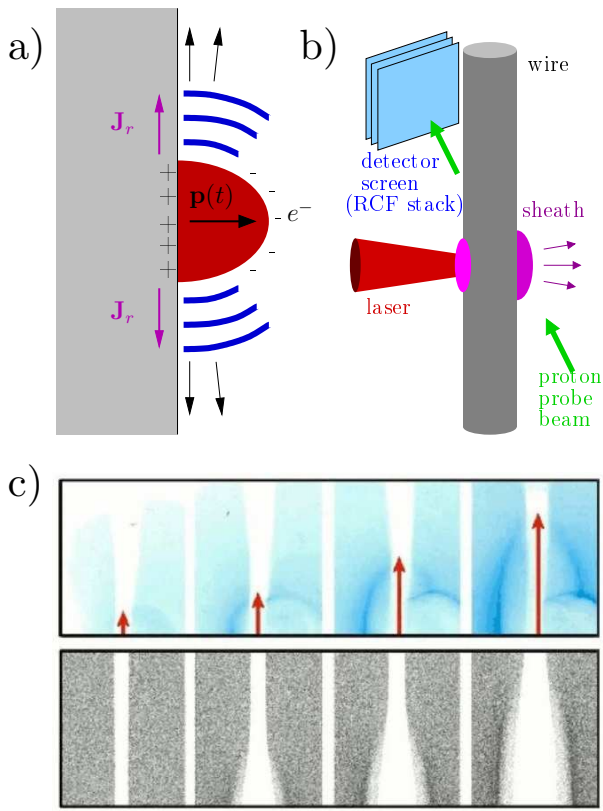}
\caption{Generation and observation of laser-driven unipolar SPs. a): sketch of unipolar SP generation by the transient charge separation associated to TNSA. The time-varying dipole ${\bf p}(t)$ acts as an antenna for the SP, which drives return currents (${\bf J}_r$) from the target edges.b): experimental set-up for proton probing of the unipolar SP propagation in a wire target. Frame~c): experimental proton images\cite{quinnPRL09} at different times (top) and related particle tracing simulations (bottom), showing the propagation of a field front at a velocity close to $c$. The laser pulse (VULCAN petawatt system at Central Laser Facility, RAL, Didcot, UK) had duration of 1~ps and intensity of $3 \times 10^{19}~{\rm W cm}^{-2}$ and the target was a $125~\mu{\rm m}$ Gold wire. Frame~c) reprinted from Ref.\onlinecite{quinnPRL09}. Used by permission.}
\label{fig:quinnexp}\label{fig:unipolar}
\end{center}
\end{figure}

Since the antenna is localized on the rear surface of the target, which is a conductor (because it is either metallic or ionized by the strong fields), it also excites SPs which propagate away from the sheath region as shown in the cartoon of Fig.\ref{fig:unipolar}~a).
This dynamics has been experimentally observed using the set-up shown in Fig.\ref{fig:quinnexp}~b), where the fields propagating along a wire target\cite{quinnPRL09} are detected using the proton probing technique\cite{borghesiPoP02}. 
Protons directed towards the sheath region are deflected by the electric field and, in a time-of-flight configuration, produce images at different times in a stack of radiochromic film (RCF) in which each layer is mostly sensitive to a given value proton energy, hence to a different probing time.  
The images (Fig.\ref{fig:quinnexp}~c) show that fields are localized near to the surface and propagate at a velocity $v=(0.95\pm 0.05)c$ as an unipolar pulse. The latter carries a net current with peak value pf $\sim 10^4$~A and is associated to transient fields exceeding $10^{11}~{\rm V cm}^{-1}$. 
Since a fraction of the fast electrons is able to escape in vacuum, the phenomenon can be described as the transient neutralization of the conductor: the SP fields produce the surface return currents which restore charge neutrality.  

\begin{figure*}
\begin{center}
\includegraphics[width=0.8\textwidth]{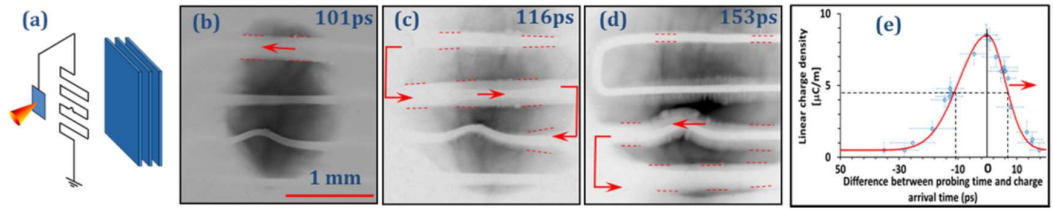}
\caption{a): set-up of a ``self-probing'' target for the study of the propagation of the unipolar SP along a wire with multiple bends\cite{karNC16}. Differently from the set-up in Fig.\ref{fig:quinnexp}~a), both the unipolar SP and the probe beam originate from the same target. b)-d): images of the wire at different times, showing the propagation of the unipolar SP. e): reconstruction of the SP pulse profile. The experiment was performed using $30~fs$ duration, $a_L \sim 10$ amplitude laser pulses from the ARCTURUS system at Heinrich Heine University in D\"usseldorf (Germany). Reprinted from Ref.\onlinecite{karNC16} with permission.}
\label{fig:coilexp-1}
\end{center}
\end{figure*}

More recent measurements have shown that the unipolar SP can propagate over bent wires for $\sim$mm distance without substantial extinction or dispersion. These latter observations have been obtained in a ``self-probing target'' arrangement, in which a single target produces both the probe protons and the unipolar pulse, as shown in Fig.\ref{fig:coilexp-1}. In this arrangement, the wire is attached is oriented in order than the SP remains in the probe field of view during the propagation. The unipolar pulse (produced by a laser system of much shorter duration than in Ref.\onlinecite{quinnPRL09}) had velocity $v=(0.96\pm 0.04)c$, peak fields $\sim 10^{11}~{\rm V cm}^{-1}$, a few picosecond duration and carried a net current of some~kA. If the wire is instead arranged as a coil with its axis perpendicular to the target surface, the fields of the traveling SP interact with the protons propagating inside the coil. With proper timing, the SP field can accelerate and focus a portion of the protons. This effect has been exploited to develop a novel device for ion post-acceleration, which leads to chromatic focusing and energy enhancement of the proton beam\cite{karNC16,ahmedSR17,kar-patent}.

\section{Conclusions}

Surface plasmons of high field amplitude have been shown to play an important role in several phenomena and applications in the context of high intensity interactions with solid targets. For instance the degree of laser absorption, the efficiency of the generation of ``fast'' electrons and the development of instabilities may be strongly affected by the excitation of surface plasmons. The experiments with high contrast pulses and grating targets have demonstrated the surface plasmon-enhancement of different types of high-energy emission, which may contribute substantially to the development of ultrashort laser-driven sources. To this aim, the very recent observations of additional enhancement effects either by static engineering of the target profile or by dynamic modification via short laser prepulses open a route for source optimization and improved control at the sub-micrometric and femtosecond scales. Finally, unipolar surface plasmons with picosecond duration have been exploited in engineered targets for post-acceleration of laser-accelerated protons, with further developments possible also for the related generation of intense THz pulses. These studies demonstrate the importance and potential of surface plasmon physics for the applications of superintense laser systems and also claim for improved theories of surface plasmons in the regime of relativistic electrons. We therefore expect high field plasmonics to further emerge as an important and innovative area of the field of high-intensity laser-matter interaction.

\begin{acknowledgments}
The author acknowledges M.~Borghesi, G.~Cantono, T.~Ceccotti, F.~Cornolti, S.~Kar, L.~Fedeli, T.~V.~Liseykina, C.~Riconda, F.~Pegoraro, A.~Sgattoni, K.~Quinn for their leading role in proposing and stimulating research on high field plasmonics and/or obtaining, analyzing and interpreting the results shown in this review in the framework of our collaborations. The contributions of H.~Ahmed, B.~Aurand, F.~Baffigi, A.~Bigongiari, M.~Bougeard, S.~Brauckmann, D.~C.~Carroll, C.~A.~Cecchetti, M.~Cerchez, R.~J.~Clarke, P.~D'Oliveira, V.~Floquet, J.~Fuchs, P.~Gallegos, D.~Garzella, L.~A.~Gizzi, P.~Hadjisolomou, A.~Heron, O.~Klimo, M.~Kv\v{e}to\v{n}, L.~Labate, L.~Lancia, C.~Lewis, P.~Martin, P.~McKenna, D.~Neely,  G.~Nersisyan, M.~Notley, F.~Novotny, M.~Passoni, A.~Pipahl, M.~Possolt, R.~Prasad, I.~Prencipe, J.~Prokupek, J.~Proska, J.~Psikal, M.~N.~Quinn, B.~Ramakrishna, M.~Raynaud, F.~Reau, A.~P.~L.~Robinson, L.~Romagnani, H.~Ruhl, G.~Sarri,  A.-M.~Scrhoer, S.~Sinigardi, L.~Stolcova, M.~Swantusch, O.~Tcherbakoff, T.~Toncian, L.~Vassura, A.~Velyhan, V.~A.~Vhsivkov,  O.~Willi, P.~A.~Wilson, X.~H.~Yuan, M.~Zepf are also gratefully acknowledged.
\end{acknowledgments}


%

\end{document}